\documentclass[%
 reprint,twocolumn,floatfix,
 amsmath,amssymb,
 aps,
 prl,
 showpacs,
longbibliography,
superscriptaddress
]{revtex4-1}

\usepackage{graphicx}
\usepackage{dcolumn}
\usepackage{bm}
\usepackage{color}
\usepackage{comment}
\usepackage{url}
\usepackage[normalem]{ulem}
\usepackage[version=3]{mhchem} 

\usepackage{amsmath}
\usepackage{amssymb}
\usepackage{lipsum}
\usepackage{url}
\usepackage{breakurl}
\usepackage[breaklinks]{hyperref}
\usepackage{comment}
\newcommand{\figurewidth}{.48\textwidth}
\renewcommand{\epsilon}{\varepsilon}

\newcommand{\tn}[1]{\textnormal{#1}}  

\newcommand*{\pH}{\tn{pH}}

\newcommand*{\pKa}{\tn{p}K_{\tn a}}

\begin{document}

\title{Collapse/expansion dynamics and actuation of pH-responsive nanogels}

  \author{Jiaxing Yuan}
\affiliation {Research Center for Advanced Science and Technology, University
  of Tokyo, 4-6-1 Komaba, Meguro-ku, Tokyo 153-8904, Japan}

\author{Tine Curk}
\email{tcurk@jhu.edu}
\affiliation{Department of Materials Science \& Engineering, Johns Hopkins University,
  Baltimore, Maryland 21218, USA}
  
\keywords{Charge regulation, electrostatic interaction, polyelectrolyte
  hydrogel, discontinuous transition}



\begin{abstract}
  Polyelectrolyte (PE) hydrogels can dynamically respond to external stimuli, such as changes in pH and temperature, which benefits their use for smart materials and nanodevices with tunable properties. 
  We investigate equilibrium 
  conformations and phase transition dynamics of pH-responsive nanogels 
  using hybrid molecular dynamics/Monte Carlo simulations with full consideration of electrostatic and hydrodynamic interactions.
  We demonstrate that PE nanogels exhibit a closed-loop phase behavior with a discontinuous swelling--collapse transition that occurs only at intermediate pH values. 
  A 50~nm nanogel particle close to a critical point functions as a pH-driven actuator with a microsecond conformational response and work density $\approx10^5~\tn{J/m}^3$, an order of magnitude larger than skeletal muscles.
   The collapse/expansion time scales as $L^{2}$ and the power density scales as $L^{-2}$ where $L$ is the linear size of the gel.
 Our work provides fundamental insight into phase behavior and non-equilibrium dynamics of the swelling--collapse transition, and our method enables the investigation of charge--structure--hydrodynamic coupling in soft materials.

\end{abstract}
\date{\today}

\maketitle

Polyelectrolyte (PE) hydrogels are three-dimensional networks of cross-linked
charged polymers that have found widespread industrial
applications~\cite{karg2019}. The conformational response of PE gels to
external stimuli, such as variation in
temperature~\cite{vanderLinden2013,sijia2021},
pH~\cite{piyush2002,glazer2012,han2020}, salt concentration~\cite{shao2021},
and electrical or magnetic fields~\cite{schreyer2000},
makes them prime building blocks for the design of smart materials and
nanodevices such as drug delivery vehicles~\cite{piyush2002}, and actuators for soft robotics~\cite{morales2014,li2020}.
Through the incorporation of stimuli-responsive components such as ionizable groups, hydrogels undergo shape changes and perform mechanical work in response to triggers such as changes in temperature and pH~\cite{li2022soft}.
Understanding both the thermodynamics and kinetics of the swelling--collapse transition of hydrogels is crucial for improving our ability to precisely control their shape change and actuation. 

A polymer gel typically undergoes a swelling--collapse transition upon a change in
temperature. This transition is continuous for a free-standing uncharged gel~\cite{quesada2011}, but can become discontinuous for gels under tension~\cite{dusek1968} or highly charged PE gels in low-dielectric solvents~\cite{tanaka1980,yan2003}. 
Moreover, hydrogels are typically weak acids/bases where the charge of the hydrogel is not constant but depends on the gel conformation due to the charge regulation (CR) effect~\cite{curk2021,curk2022}. 
Charge regulation leads to a discontinuous swelling--collapse transition of pH-responsive hydrogels, which terminates in a critical point at a finite pH value, as is predicted by mean-field theory~\cite{tanaka1980,polotsky2013,zheng2023}. 
While extensive research has been conducted on the phase diagram of bulk PE hydrogels~\cite{tanaka1980,polotsky2013,yan2003,zheng2023}, there is a notable lack of attention given to the swelling--collapse kinetics. This is especially the case for nanogels since conventional microscopic techniques struggle to capture the time evolution of nanoscale conformations~\cite{keidel2018time,dallari2024real}. In particular, nanogels possess substantial surface-to-volume ratio~\cite{zhou2020recent,wang2021strategic}
and offer faster dynamical response to the external stimuli~\cite{asadian2012functional}, making them highly promising for the emerging biomedical and nanotechnology applications.

In this Letter, we explore the charge--structure--hydrodynamic coupling in a PE nanogel. We use hybrid molecular dynamics (MD)/MC simulations to investigate equilibrium conformations and transition dynamics of a pH-responsive nanogel, focusing on the impact of electrostatic and hydrodynamic interactions on the collapse and expansion kinetics. From these we derive design guidelines for engineering pH-driven nanoactuators. 

\begin{figure}
  \centering \includegraphics[width=\figurewidth]{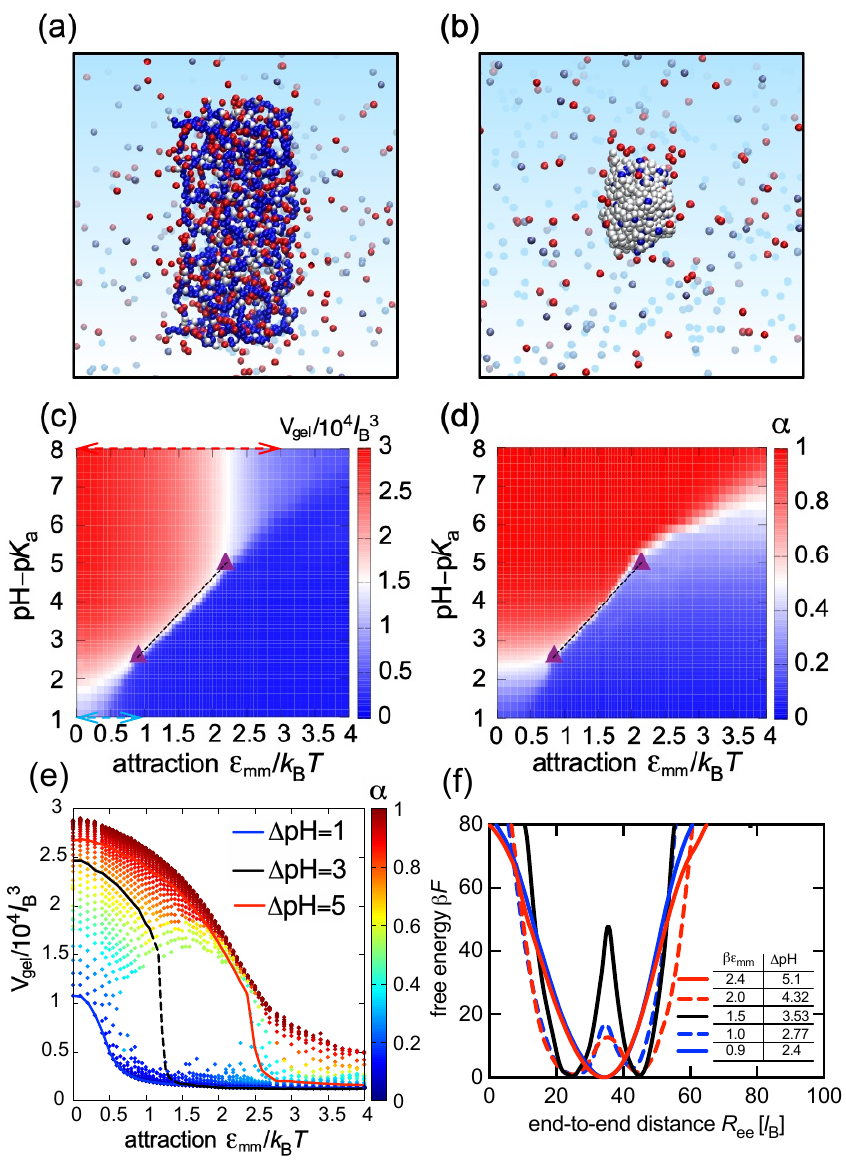}
  \caption{Equilibrium phase diagram of a nanogel. (a,b) Typical expanded ($\Delta\pH=6$) and collapsed
    ($\Delta\pH=1$) conformations at $\varepsilon_\tn{mm}=k_{\tn{B}}T$ showing negatively charged dissociated (blue) and
  neutral (white) acid groups, free cations (red), and anions (metallic blue), note $\Delta\pH=\pH-\pKa$.
    (c,d) Average gel volume~$V_\tn{gel}$ and degree of ionization~$\alpha$
    showing a sharp transition at intermediate $\Delta\pH$ and
    $\varepsilon_{\tn{mm}}$. Triangles in (c) and (d) mark the locations of
    the peaks of equilibrium fluctuations in $V_\tn{gel}$ and $\alpha$. The dashed black lines mark the discontinuous transition and the arrows represent the kinetic pathway explored below. (e) 
    Average~$\alpha$ 
    displaying a closed-loop phase diagram with a central discontinuous
    transition. (f) The transition free-energy barrier 
    along the dashed black line in (c,d). The volume~$V_{\tn {gel}}$ is obtained from the convex
    hull of the gel monomers. Parameters: width $N_x=N_y=2$ and length
    $N_z=6$ unit cells of size~$N_{\tn p}=10$, containing a total of 1395
    ionizable monomers.}
\label{fig:pd}
\end{figure}

To understand the actuation behavior, we first calculate the equilibrium phase diagram of a nanogel particle. We employ a coarse-grained PE model where all charges and all ionizable sites in the hydrogel are explicitly represented, allowing accurate calculation of electrostatic interactions.  
Moreover, the model captures
the dynamic interplay between protonation, gel conformation, electrostatic
interaction, counter-ion osmotic pressure, and excluded-volume effects.
The anionic gel is immersed in an aqueous solution of
Bjerrum length $l_{\tn{B}} = \beta q_0^2/(4\pi\varepsilon_\tn{sol})$,
where $\beta=1/(k_{\tn{B}}T)$, $k_{\tn{B}}$ is Boltzmann constant,
$T$ is the absolute temperature, $q_0$ denotes the elementary charge,
and $\varepsilon_{\text{sol}}$ is the solvent permittivity, resulting in
$l_{\tn B}=0.72\, \tn{nm}$ for a hydrogel at room temperature. The gel consists of ionizable bead-spring PE chains
 cross-linked into
a gel network with crosslinking distance $N_{\tn{p}}$ [Fig.~\ref{fig:pd}(a)--(b)].

Ions and monomers are represented as spheres of diameter
$\sigma=l_{\tn{B}}$, which is a typical size of hydrated ions or ionizable
groups.
The short-range attraction between PE monomers is represented by the
Lennard-Jones potential with strength~$\varepsilon_{\tn{mm}}$, which determines the solvent quality, and cutoff distance $r_\text{cut}=3\sigma$.
Each ionizable site (monomer) is subject to the acid dissociation
reaction, $\tn{A}^0\rightleftharpoons \tn{A}^- + \tn{H}^+$, where $\tn{A}^0$
and $\tn{A}^-$ are the neutral and dissociated states with $\tn{H}^+$ the
dissociated proton.  The probability~$\alpha_i$ that a site~$i$ at
position~$\mathbf{r}_i$ is charged (dissociated) is determined
by~\cite{podgornik2018,curk2021},
\begin{equation}
\frac{\alpha_i}{1-\alpha_i} = 10^{\pH-\pKa} e^{-\beta \psi(\mathbf{r}_i)
    q_0} \;,
\label{eq:alphai}
\end{equation}
with $\pKa$ the equilibrium dissociation constant and
$\psi(\mathbf{r}_i)$ the electrostatic potential at site~$i$. We model all charged entities explicitly 
and perform grand-canonical MC exchange of dissociated ions ($\tn H^+$)  and monovalent salt cations ($\tn S^+$) and anions ($\tn
S^-$) with a reservoir at a given $\tn{pH}$ and monovalent salt concentration $c_\tn{s}=10^{-3}$M. 
Electrostatic interactions are calculated using particle--particle particle--mesh algorithm~\cite{hockney-eastwood} with relative force accuracy $10^{-3}$. 

The system configurations are evolved employing the standard velocity-Verlet MD algorithm, while ionization
states [Eq.~\eqref{eq:alphai}] and ion exchange are sampled using an efficient charge regulation MC
(CR-MC) solver~\cite{curk2022}.  To investigate the equilibrium phase diagram, we employ the standard Langevin dynamics simulations.
The phase transition dynamics of PE are influenced by hydrodynamic interactions (HI)~\cite{Kikuchi2005,Kamata2009,Yuan2022Impact,yuan2024}.
Thus, for the study of non-equilibrium collapse/expansion dynamics of nanogels, we account for HI by combining the CR-MC solver with a method based on
dissipative particle dynamics (DPD)~\cite{grootwarren} that couples the solute (gel) particles to the DPD solvent (DPDS)~\cite{curk2024dissipative}. 
Additional simulation details are provided in Supplemental Material~\cite{Supple}.


We initially focus on an elongated nanogel which is a prototypical system for 
hydrogel-based soft actuators and artificial muscles~\cite{park2020}.  
Based on experimental observations~\cite{tanaka1980} and theoretical~\cite{polotsky2013,zheng2023} 
predictions, we anticipate that a hydrogel exhibits a critical point beyond which the collapse--swelling transition is discontinuous. 
We compute the pH--$\epsilon_\tn{mm}/k_\tn{B}T$  phase diagram of a nanogel and find a sharp collapse--swelling transition at intermediate $\pH$ and $\epsilon_\tn{mm}/k_\tn{B}T$ values [Fig.~\ref{fig:pd}(c,d)].
The discontinuous transition arises due to CR coupling between the gel conformation and ionization (Eq.~\eqref{eq:alphai}), leading to coexistence between an expanded charged gel and a collapsed uncharged gel. However, this CR coupling and the discontinuous transition disappear in the limits $\pH\to\pm\infty$ at which the charge on the gel is constant ($\alpha\to1$ or $\alpha\to0$). Fluctuation
analysis of $V_\tn{gel}$ and $\alpha$ (Fig.~S1 in \cite{Supple}) further shows two distinct peaks that indicate tentative locations
of the two critical points [marked by triangles in
Fig.~\ref{fig:pd}(c)--(d)]. The peak locations are consistent for fluctuations in gel volume~$V_\tn{gel}$ and ionization~$\alpha$. 
Thus, interestingly, the discontinuous region is limited to intermediate pH values, implying a closed-loop phase diagram [Fig.~\ref{fig:pd}(e)]. Free energy calculations [Fig.~\ref{fig:pd}(f); also see Fig.~S2 in \cite{Supple}] confirm two stable states separated by a barrier at intermediate pH. Finite-size scaling shows the barrier increases with increasing gel volume (Fig.~S3 in \cite{Supple}),  which further supports that the
conformational change is a discontinuous transition at intermediate pH values. 
On the other hand, the barrier disappears at both high and low pH values indicating that a
continuous transition occurs for both neutral (high pH, $\alpha\sim0$) and
fully ionized (low pH, $\alpha\sim1$) nanogel (Fig.~S4 in \cite{Supple}).

The location of the transition is not significantly affected by varying the shape of the gel or the crosslink distance (Fig.~S5 in \cite{Supple}), which implies that the phase diagram in Fig.~\ref{fig:pd}(c)--(d) is robust and is expected to apply to randomly cross-linked PE hydrogels. Indeed, our findings can explain why
the experimentally observed hydrogel volume transition is sharpest at intermediate temperature and
pH~\cite{Seong2001}.

\begin{figure}
\centering \includegraphics[width=\figurewidth]{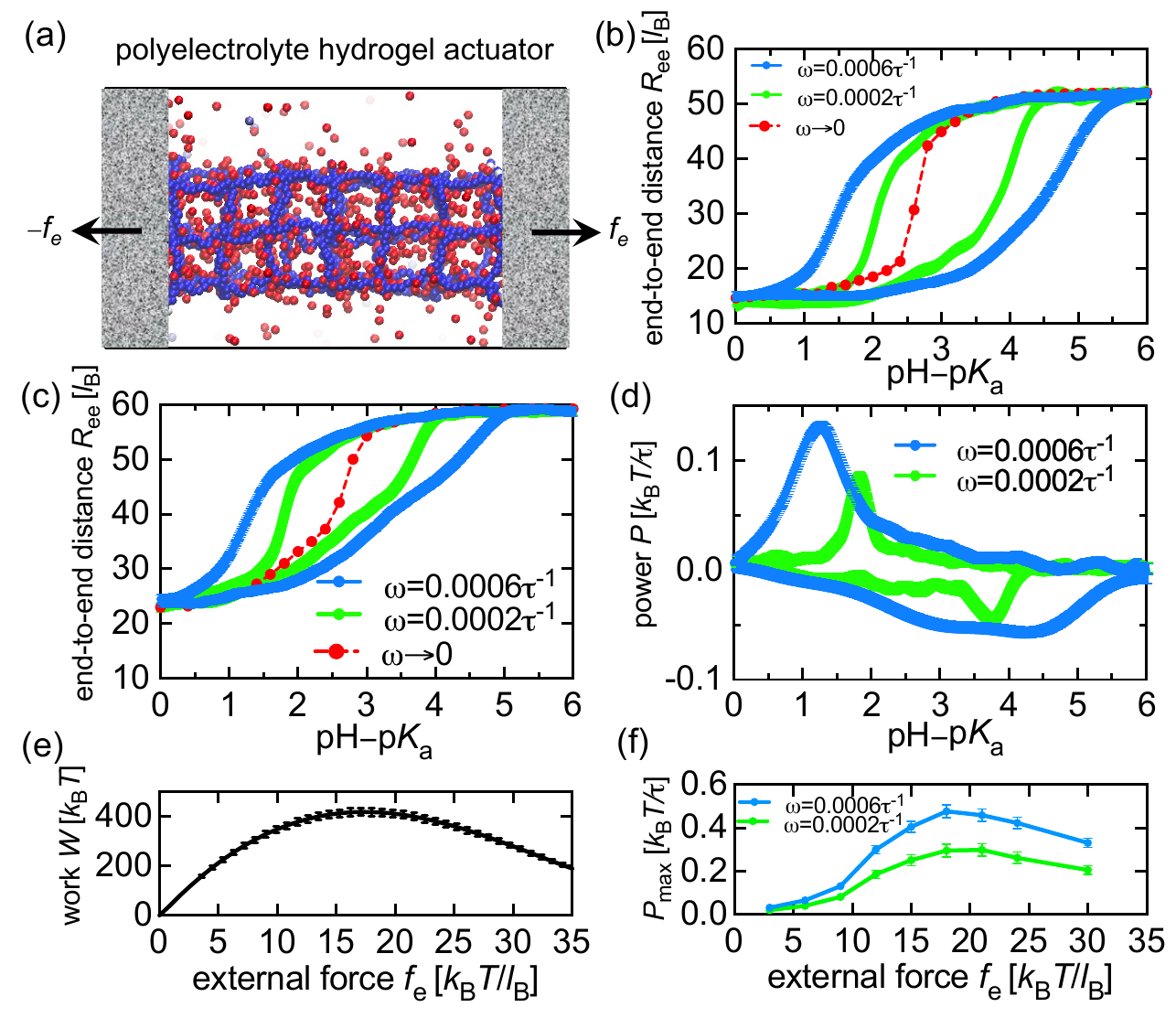}
\caption{Polyelectrolyte hydrogel as a pH-responsive soft actuator. (a) An
  external force $f_\tn{e}$ is applied to the ends of the gel. (b,c) The
  response of a weak polyelectrolyte gel ($\pKa=4$) under pH ramping
  rate~$\omega$ at $f_\tn{e}=0$ (b) and $f_\tn{e} = 9 k_{\tn{B}}T/l_{\tn B}$
  (c).  (d) The corresponding power output, $P=-f_\tn{e}dR_\tn{ee}/dt$, at
  $f_\tn{e} = 9 k_{\tn{B}}T/l_{\tn B}$.  (e) The work performed 
  against the external force upon changing $\pH$ from $\pH_{1}=\pKa+6$ to
  $\pH_{0}=\pKa$ under $w\to 0$. (f) The maximum power~$P_\text{max}$ as a function of external force
  at the ramping rate $\omega=0.0006\tau^{-1}$ and $\omega=0.0002\tau^{-1}$.
  The error bars show standard errors.}
\label{fig:pH_ramp}
\end{figure}

Having calculated the equilibrium phase diagram of the nanogel, we turn to kinetics and design a
pH-responsive actuator. We introduce a constant force with magnitude
$f_\tn{e}$ to the two ends of the hydrogel [Fig.~\ref{fig:pH_ramp}(a)] and apply
a zigzag time-varying pH cycle between $\pH_\tn{0}=\pKa$ and
$\pH_{1} = \pKa+6$ with ramp rate~$\omega={d(\tn{pH})}/{dt}$ (see \cite{Supple} for details).
The total time to complete a full cycle is $t_\tn{f} = 2(\tn{pH}_\tn{1}-\tn{pH}_\tn{0})/\omega$.  
Guided by the phase diagram [Fig.~\ref{fig:pd}(c)--(d)] we chose
$\varepsilon_{\tn{mm}}=1k_{\tn B}T$, which is close to the peak in equilibrium fluctuations and is thus expected to result
in a sharp and fast response without exhibiting strong hysteresis.
Hydrodynamics is resolved using the DPDS method~\cite{curk2024dissipative} that models an aqueous solution with time unit $\tau=0.029~\tn{ns}$, while the ionization kinetics is modeled using the CR-MC method~\cite{curk2022} with realistic ionization rates $k_{\tn d} \approx 3 \times 10^{6}~\tn{s}^{-1}$~\cite{Kanzaki2014}. 

We find that the end-to-end distance~$R_\tn{ee}$ of the gel follows the pH
cycle and shows hysteresis that
becomes more pronounced at faster cycling rates [Fig.~\ref{fig:pH_ramp}(b)].  The response under external
force [Fig.~\ref{fig:pH_ramp}(c)] remains qualitatively the same, but
conformations are more elongated due to the stretching by the external force. 
The maximum instantaneous power is $P_\tn{max} \approx 0.1 k_{\tn{B}}T/\tau$
at the high-frequency limit ($\omega=0.0006\tau^{-1}$) allowable by protonation dynamics~\cite{Kanzaki2014} [Fig.~\ref{fig:pH_ramp}(d)].
The response time of the gel at lower rate $\omega=0.0002\tau^{-1}$ is $\tau_{\tn r} \approx 3/\omega \approx0.4~\mu\tn{s}$, using the DPD timescale $\tau\approx0.029$~ns~\cite{curk2024dissipative}, which is at least 4 orders of magnitude faster
than typical values for mammalian skeletal muscles ~\cite{ranatunga1998,tissaphern2007}
or macroscopic soft actuators~\cite{Kim2015,Ma2020}. This fast response is a
consequence of the nanoscale width of the gel which permits fast transport of solvent.

The work~$W$ performed during the contraction of the gel is
$W = f_{\tn{e}}\Delta R_\mathrm{ee}$, with the change in the end to end distance $\Delta R_\mathrm{ee}= [R_{\tn{ee}}(\pH_1) - R_{\tn{ee}}(\pH_0)]$. $W$ reaches a maximum
at force $f_\tn{e}^*\approx 18 k_{\tn B}T/l_{\tn B}$ and decreases at larger forces
that prevent the gel from fully collapsing [Fig.~\ref{fig:pH_ramp}(e)]. 
The optimal load translates to stress
$\sigma_{f} = f_{\tn{e}}^* / A \approx 5\times10^5~\tn{Pa}$, with the cross-section area,
$A\approx N_{\tn p}^2 N_x N_y \sigma^2 = 400 l_{\tn B}^2$, under which conditions the gel can
elongate twofold. Moreover, the maximum work~$W^*$ at the force~$f_{\tn{e}}^*$ is $W^*\approx400k_\tn{B}T$ [Fig.~\ref{fig:pH_ramp}(e)], corresponding to a work density of $W^*/V_\tn{gel}\approx150\tn{kJ}/\tn{m}^3$.
Thus, the maximum strain of the pH-driven gel actuator is comparable to that of skeletal muscles, while the stress and work density are an order of magnitude larger~\cite{tissaphern2007}.
The maximum instantaneous power $P_\tn{max}$ at the high frequency limit ($\omega=0.0006\tau^{-1}$) also exhibits a peak close to $f_\tn{e}^*$ 
[Fig.~\ref{fig:pH_ramp}(f)], suggesting that the nanogel actuator
has the potential to yield favorable performance in terms of both work and power even at fast cycling rates.

\begin{figure}
\centering \includegraphics[width=\figurewidth]{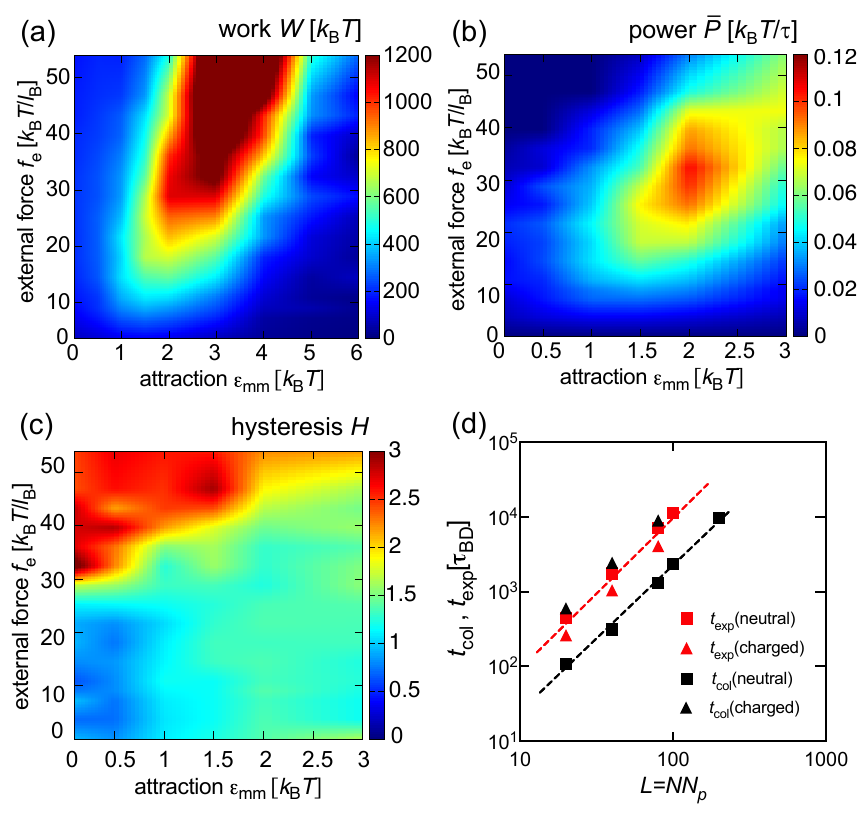}
\caption{Performance of polyelectrolyte nanogel actuator. (a) Equilibrium work per contraction~$W=f_\tn{e}\Delta R_\mathrm{ee}(\omega \to 0)$, (b) mean power~$\bar{P}$, and (c) hysteresis $H$ (see Eq.~\ref{eq:width}) at ramping rate $\omega = 0.0006/\tau$. 
(d) The collapse timescale~$t_\text{col}$ and the expansion timescale~$t_\text{exp}$ of a neutral and fully charged gel as a function of gel size~$L$.  The dashed lines have a slope of 2.}
\label{fig:work_power}
\end{figure}

To explore the optimal design of the actuator we calculate the work, power, and hysteresis over a full spectrum of $\epsilon_{\tn{mm}}/k_{\tn B}T$ and $f_{\tn{e}}$ (Fig.~\ref{fig:work_power}). The equilibrium work performed per contraction, $W(\omega\to0)$, 
is largest in the discontinuous region of the phase diagram and strong forces (under external force, the discontinuous region moves to larger $\epsilon_\tn{mm}$) [Fig.~\ref{fig:work_power}(a)]. At fast cycling rates, however, the response is limited by the viscous drag and we find the mean power per contraction, 
\begin{equation}
\bar{P}(\omega)=W(\omega) \frac{\omega}{\pH_1-\pH_0} \;,
\end{equation}
 is maximal at intermediate $f_{\tn{e}}$ and $\epsilon_{\tn{mm}}$ [Fig.~\ref{fig:work_power}(b)]. 
This observation underscores the crucial role of hydrodynamic effects: the slow relaxation dynamics at $\epsilon_{\text{mm}}/k_{\text{B}}T > 2$ fail to keep pace with the pH ramping rate. Note that the power plot is obtained at the high frequency limit allowable by protonation dynamics, $\omega=0.0006/\tau$~\cite{Kanzaki2014}, thus the work and power plots in Figs.~\ref{fig:work_power}(a)--\ref{fig:work_power}(b) effectively show the low and high frequency limits, respectively.

To quantify the hysteresis we calculate the effective width~$H$ between the forward and reverse process,
\begin{equation}
H(\omega)= \frac{1}{\Delta R_\tn{ee}(\omega)} \oint R_\mathrm{ee}  d (\pH) 
\label{eq:width}
\end{equation}
 as the integration area of the closed loop in $R_\mathrm{ee}$--$\pH$ divided by the total change in size $\Delta R_\mathrm{ee}(\omega)$ [Fig.~\ref{fig:work_power}(c)]. Hysteresis is lowest at $\epsilon_\tn{mm}/k_\tn{B}T\le1$, which coincides with the continuous region of the phase diagram [cf. Fig.~\ref{fig:pd}(c)]. The hysteresis effect becomes significantly stronger under a faster ramping rate of $\omega=0.003\tau^{-1}$ [Fig.~S6(c) in~\cite{Supple}] at which ionization equilibrium cannot be achieved 
($k_\tn{d} (\tn{pH}_\tn{1}-\tn{pH}_\tn{0})/\omega < 1$).

These findings indicate that the overall optimal condition for actuator performance is close to the critical point, $\epsilon_{\text{mm}}/k_{\text{B}}T=1$, resulting in fast shape changes with appreciable power output, but without strong hysteresis.

As the timescales of collapse~$t_\tn{col}$ and expansion~$t_\tn{exp}$ directly determine the maximum power, it is also essential to understand the impact of nanogel size~$L$ on the swelling--collapse dynamics. 
 We consider a cubic nanogel where $N_x=N_y=N_z=N$ at crosslinking distance $N_{\tn p}=10$ and calculate the expansion/collapse timescales as a function of size $L=N N_{\tn p}$.
 The red and blue arrows in Fig.~\ref{fig:pd}(c) denote the pathway of four different cases: the collapse or expansion of fully charged and neutral gels (see \cite{Supple} for the detailed protocols to induce collapse/expansion dynamics).

Interestingly, we find that both the collapse and expansion timescales exhibit scaling with the gel size~$L$ as $t_\text{col}\sim L^2$ and $t_\text{exp}\sim L^2$ [Fig.~\ref{fig:work_power}(d)] for both 
charged and neutral gels (see Fig.~S7 and \cite{Supple} for details on the determination of timescales). This aligns with a theoretical prediction for a swelling timescale of a neutral gel~\cite{tanaka1979kinetics}. Moreover, the expansion of a neutral gel is much slower than its collapse [Fig.~\ref{fig:work_power}(d)], which is consistent with experimental data~\cite{dallari2024real}. Given the work $W\sim L^3$ (the force is proportional to $L^2$ and the change in length to $L$), it follows that the power density of the actuator scales as $P/V_\tn{gel}\sim L^{-2}\sim V_\tn{gel}^{-2/3}$. For the simulated nanogel actuator [Fig.~\ref{fig:pH_ramp}(f)], the peak power is $0.5k_\tn{B}T/\tau\approx8\times10^{-11}$W and the gel volume is $V_\text{gel}\approx N_xN_yN_z N_\tn{p}^3\sigma^3\approx 10^{-5}\mu\text{m}^3$, thus the maximal power density is $P_\tn{max}/V_\tn{gel} \approx 10^{-5} W/\mu\text{m}^3$. Using the scaling we predict that for a microgel particle of $V_\text{gel}=10^3\mu\text{m}^3$, the power density is about $0.04\text{W/mm}^3$.
This showcases remarkable performance compared to the current hydrogel actuators~\cite{park2024hydrogels}. For instance, a bioinspired strong contractile hydrogel with a similar volume of $\sim10^3\mu\text{m}^3$ demonstrates a power density of 
$\sim5\times10^{-8}\text{W/mm}^3$~\cite{ma2020bioinspired}.

We evaluate the effects of ion diffusion and protonation kinetics on the actuator performance in Appendix~\ref{sec:AppendixA}. Our model does not include effects related to orientational ordering of water and related changes of permittivity close to ions, we discuss these aspects in Appendix~\ref{sec:AppendixB}.

  In summary, by dynamically modeling the charge--structure coupling via charge regulation of individual
  dissociable groups, we have calculated the pH--$\epsilon_\tn{mm}/T$ and $V_\tn{gel}$--$\epsilon_\tn{mm}/T$ phase diagrams of a nanogel particle. We demonstrate a discontinuous swelling--collapse transition for intermediate values of pH and interaction strength~$\epsilon_\tn{mm}/T$. The phase diagrams serve as a guide for predicting and engineering nanogel materials. 
By modeling the dynamics of collapse/expansion, including full hydrodynamic interactions, we demonstrate that the charge--structure--hydrodynamic coupling enables the gel to serve as an effective pH-responsive nanoactuator; the maximum strain of the proton-driven gel actuator is comparable to that of skeletal muscles, while the tensile stress and work density are an order of magnitude larger.

  Crucially, we show that the optimal performance in terms of high power and low hysteresis is found near a critical point, $\varepsilon_\tn{mm}\approx k_{\tn{B}}T$, while the power density decreases with the gel volume as $P/V_\tn{gel} \sim V_\tn{gel}^{-2/3}$.
  These scaling results show that the response of bulk hydrogels is very slow due to solvent diffusion. We hypothesize that a fast response in a bulk material could be achieved by introducing holes in a bulk gel through which solvent could be pumped or by stacking individual thin strips of the nanogel together with a proton source (e.g. electrodes) akin to the structure of biological muscles. Such a hierarchical structure could attain a fast, kHz response in a macroscopic gel.
Our Letter reveals the dynamic interplay among conformation, solvent flow, and ionization in  the swelling--collapse transition of nanogels. Our simulation setup enables the investigation of charge--structure--hydrodynamic coupling 
and can be utilized to investigate a wide range of solvated systems, including polyelectrolytes, proteins, and soft materials.

\begin{acknowledgments}
This work was supported by the startup funds provided by the Whiting School of Engineering at JHU. 
The simulations were partially performed using the Advanced Research Computing at Hopkins (rockfish.jhu.edu), which is supported by the National Science Foundation (NSF) grant number OAC 1920103. We thank Erik Luijten, Hajime Tanaka, James D. Farrell, and Michael Falk for discussions and comments on the manuscript.
\end{acknowledgments}

\appendix
\vspace{5mm}
\centerline{\bf Appendix}

\section{Effects of ion diffusion and protonation kinetics}
\label{sec:AppendixA}

Here we evaluate the effects of ion diffusion and protonation kinetics on the actuator performance. 
The timescale for ion diffusion across the nanogel studied is $t_\tn{diff}=L^2/(2D)$, which, for typical ion diffusivity $D=\tn{nm}^2/\tn{ns}$, is at least an order of magnitude faster than the fastest ramping rates considered in Figs.~\ref{fig:pH_ramp} and~\ref{fig:work_power} ($L_x/\sigma=N_xN_\tn{p}=20$). Since the timescales of collapse/expansion [Fig.~\ref{fig:work_power}(d)] and ion diffusion have the same scaling with the gel size ($\sim L^2$), we predict that ion diffusion across the gel is not expected to be a limiting process, even for bulk gels.
The protonation dynamics is governed by typical weak acid dissociation rate
$k_{\tn d} \approx 3 \times 10^{6}~\tn{s}^{-1}$ and association rate
$k_{\tn{a}} \approx 10^{11-\pH}~\tn{s}^{-1}$~\cite{Kanzaki2014}, which do not depend on the gel size. Since the response time due to solvent drag scales as $L^2$, we conclude that the protonation dynamics can be a limiting factor only for nanoscale gels ($L\lesssim 50~\tn{nm}$) or extremely weak PEs where collapse occurs at high pH. For example, the actuation power in Figs.~\ref{fig:pH_ramp}d and~\ref{fig:work_power}b are obtained at the high frequency limit allowable by protonation dynamics. The gel actuator is driven by protons and requires a proton source/sink, e.g., an electrode, and so ion/proton diffusion from the electrode to the gel could be another limiting factor.  To achieve fast kHz or even MHz response time, the proton/source sink should be placed next to the gel, and it could itself be a soft electrode, e.g. a conjugated polyelectrolyte. 
\vspace{5mm}
\section{Model limitations}
\label{sec:AppendixB}
Our model does not capture the orientational ordering of water near ions~\cite{van2016water,shi2023impact}, which can enhance electrostatic coupling at small ion--ion distances.  
Moreover, 
ion and water diffusion are generally enhanced in polarizable compared to non-polarizable models~\cite{nguyen2018influence}. 
Thus, water ordering and polarizability are anticipated to enhance the transport and the response of charge to conformational changes, potentially improving the actuator's performance.


%

\end{document}